\documentclass[preprint,nofootinbib,aps,superscriptaddress,showpacs]{revtex4}
\usepackage{color,graphicx,shortvrb,epsfig}

\usepackage{graphicx}
\usepackage{dcolumn}
\usepackage{bm}

\usepackage{colordvi}
\usepackage{psfig,epsfig,amssymb,amsmath,latexsym}

\newcommand{\no}{\nonumber}
\def\lsim{\mathrel{\rlap{\lower4pt\hbox{\hskip1pt$\sim$}}
    \raise1pt\hbox{$<$}}}         
\def\gsim{\mathrel{\rlap{\lower4pt\hbox{\hskip1pt$\sim$}}
    \raise1pt\hbox{$>$}}}         

\def\beq{\begin{equation}}
\def\eeq{\end{equation}}
\def\ba{\begin{eqnarray}}
\def\ea{\end{eqnarray}}

\def\<{\langle}
\def\>{\rangle}

\begin{document}

\begin{flushright}
\date{}
\end{flushright}

\title{Comment on ``Writhe formulas and antipodal points in plectonemic 
DNA configurations''}
\author{Joseph Samuel} 
\author{Supurna Sinha}
\address{Raman Research Institute, Bangalore, India}
\author{Abhijit Ghosh}
\address{Max Planck 
Institute for Polymer Research, 
10 Ackermannweg, 55128 Mainz, Germany
}
\begin{abstract}
We point out that the disagreement between the paper by Neukirch and 
Starostin (Ref.[1]) and 
ours (Ref. 
[5]) is only apparent and stems from a difference in approach.
Ref. [1] is concerned with classical elasticity and individual curves
while Ref. [5] focuses on statistical averages over curves.  
\end{abstract}
\maketitle

Recently Neukirch and Starostin \cite{neukirch} have noted 
that 
the use of Fuller's formula for writhe may not always be justified 
in 
analysing  experiments \cite{strick} 
that stretch and twist 
single DNA molecules. 
They have criticised some earlier works
[17-34 of Ref.\cite{neukirch}] for using Fuller's formula
\cite{F.BrockFuller041971,F.BrockFuller081978} without
always being careful to check the
conditions for its validity. Readers of this 
paper\cite{neukirch} may derive the impression
from some remarks made about \cite{sam},
that Ref.\cite{neukirch} somehow
invalidates the conclusions of our paper \cite{sam}, 
which shows how 
Fuller's formula {\it can} be used in understanding entropic DNA elasticity. 
Here we note that the apparent differences can be traced to a difference
in approach to the problem. 
We will set the comment in perspective by 
explaining the two points of view that have influenced the 
literature on bio-polymer elasticity.
We summarise the discussion by concluding that the remarks
made in \cite{neukirch} in no way invalidate any of the claims made
in \cite{sam}.

Two communities with slightly different approaches have been working on 
bio-polymer elasticity. One point of view (See, 
for instance, \cite{PhysRevLett.93.198107})
is purely mechanistic and has its roots in classical
elasticity. While this point of view captures some of the qualitative
features of single molecule experiments 
and works well in the energy dominated regime of stiffer
polymers like Actin and Microtubules, it fails to capture the regime where 
there is a competition
between the intrinsic elastic energy of the polymer
and thermal fluctuations which are present in a real cellular environment.
This latter regime which is better explained from a
statistical mechanical point of view (See, 
for instance, \cite{mezardPhysRevLett.80.1556}),
where the central notion is the partition function of the system.
These two communities view bio-polymers from slightly different 
perspectives: the mechanistic view emphasizes individual configurations, 
while the statistical view averages over configurations and 
focuses on the partition function, which is related to experimentally 
accessible quantities.  
The differences between our paper (Ref[5]) and Neukirch
and Starostin's paper (Ref[1]) 
comes from these two distinct
viewpoints.

The view offered in \cite{sam}, is that while it is incorrect to claim
that the partition function of the $SAWLC$ model 
{\it equals} that of the worm like rod chain $WLRC$ 
{\it exactly} (as 
\cite{mezardPhysRevLett.80.1556} 
appear to do),
it is nevertheless a good {\it approximation} over a range of forces and torques.
To see this, note that 
changes in the two notions of writhe (Fuller Writhe $W_{F}$ and 
C\u{a}lug\u{a}reanu-White
writhe $W_{CW}$) are equal
($\delta W_{CW}({\cal C})=\delta W_{F}({\cal C})$) 
for small variations
of the curve ${\cal C}$ (provided both quantities are well
defined in the variation). Integrating
this equation we find that the difference $W_{CW}({\cal C})-W_{F}({\cal 
C})$
is constant for deformations of ${\cal C}$ which are
neither self crossing nor south crossing.
We will follow \cite{neukirch} in referring to  
such deformations 
as ``good'' deformations. 
We choose for a reference curve the straight
line in the $z$ direction  
(${\hat t}={\hat z}$).
Noting that the constant vanishes on the reference curve, we
arrive at Fuller's formula
$W_{F}=W_{CW}$ for all curves which can be deformed to the reference
curve by ``good'' deformations. 
We refer to curves related to the reference curve by ``good'' deformations
as ``good'' curves. Note that the set of ``good'' curves
is much larger than just small perturbations about the straight line.
For instance, curves which are nowhere back-bending ($t_z\ge0$)
are ``good'' curves and these may be far from straight.

The main point we make in Ref [5] is that for a range of $(F,W)$,
the set of ``good'' curves dominates both partition functions and as a result, 
$Z_{SAWLC}(F,W)$ 
equals
$Z_{WLRC}(F,W)$  
{\it approximately}.
\begin{equation}
Z_{SAWLC}(F,W)\approx Z_{WLRC}(F,W)
\label{approx}
\end{equation}
The accuracy of the approximation is determined by the extent to which
the ``good'' curves dominate the partition function.
A simple example clarifies the matter.
Suppose we wish to find the expectation of the function $h(x)$:
$h(x)=x^2$  when $-3 <x <3$; $h(x)=-1$ otherwise,
defined piecewise over the real
line with a Gaussian measure $exp(-\alpha x^2)$, suitably normalised. 
Consider the function
$g(x)=x^2$ everywhere.
Certainly $h(x)$ is not equal to $g(x)$, as there are points where they 
differ considerably. However, in computing the expectation 
value
$<h(x)>$ of $h(x)$
one can approximately replace it by $<g(x)>$ which is
analytically more tractable.
$<h(x)>$ is approximately equal to $<g(x)>$.
The approximation is good if $\alpha$ is not too small.
Likewise, for forces which are not too small, the approximate partition 
function
is expected to be close to the exact one. This explains the
efficacy of Bouchiat and Mezard's WLRC model
beyond perturbation theory\cite{nelson,sinha}.

Both Ref.\cite{neukirch} and Ref.\cite{sam}
attempt to understand the two writhe formulae and
their use in the non perturbative regime.

However, the emphasis in \cite{neukirch} is on individual 
configurations and not in the statistical sense.
It is indeed true as both Ref. [1] and Ref. [5] note,
that the writhe formulae are {\it not} the same on individual 
curves. 
Our claims of approximate equality of the two models
are made at the level 
of the partition function and not individual curves.
We claim that the partition functions  
are approximately equal (Eq. (\ref{approx})) to 
each other for a wider range of parameters than one would naively 
expect.   
Such an approximation works very well
at very high forces. 
Let us now consider twisting the molecule:

(a). {\it At low twist:} This is the paraxial limit where the backbone of 
the polymer is essentially straight and 
the tangent vector ${\hat t}$ to the polymer explores
the neighborhood of the north pole \cite{nelson,sinha}. 
This is the perturbative regime, which 
is not controversial.

(b). {\it At intermediate twist:} In this regime we have a writhing 
polymer which may not be nearly straight. However, our argument 
that ``good'' curves dominate the partition function applies 
and we conclude that Eq. (\ref{approx}) holds.

(c). {\it At high twist:} The energy cost of accommodating writhe
becomes nearly zero in self avoiding models. This is because an 
infinitely thin polymer can writhe at negligible energy cost 
by winding around itself as a plectoneme. (The word is Greek
for ``twisted thread'' and describes the structures often seen on
telephone cords.) To render the energy finite, one has to 
``fatten'' the thread and allow for
the finite thickness of the DNA molecule (about $2\rm{nm}$). 
This pathology of infinitely thin threads is well known to 
mathematicians and was quite early noticed by Fuller 
\cite{F.BrockFuller041971,F.BrockFuller081978}. 

A similar pathology also afflicts south avoiding models:
writhe can be stored at negligible energy cost by winding around
the south pole. These configurations can be described, 
mixing our small 
Latin and less Greek,
as ``Australonemes'' (southerly threads). If 
one excludes
a finite region around the south pole (as Bouchiat and Mezard do 
\cite{mezardPhysRevLett.80.1556}) by using a cutoff,
one ends up with the same finite energy cost per unit writhe for 
appropriate cutoff.
As a result, even in the high twist limit one finds that 
the partition functions are approximately equal
and Eq. (\ref{approx}) holds.

To summarize, we have clarified the issues
surrounding the use of Fuller's formula for the writhe of a space curve.
We note that the formula, {\it when used with due care}, can be a valuable 
aid to taming an otherwise intractable calculation. It permits
an approximate determination of the partition function and a prediction
for the experimentally measured twist-extension relations in the 
presence of an applied force. We observe\cite{sam}
using the Closed Circuit theorem that self avoidance and 
south avoidance have the same topological effect of obstructing link 
release and topological untwisting. 
Our observations justify theoretical work by Bouchiat and Mezard, which 
though successful in interpreting the data, 
has been criticised for the incorrect use of Fuller's formula.
We thus understand the ``unreasonable effectiveness''
of Fuller's formula in understanding DNA elasticity.

{\it Acknowledgements:}
It is a pleasure to thank S. Neukirch for discussions.

\end{document}